\documentclass[%
 reprint,
 amsmath,amssymb,
 aps,
 prl, nolongbibliography
]{revtex4-2}

\usepackage{graphicx}
\usepackage{dcolumn}
\usepackage{bm}

\usepackage{amsmath}
\usepackage{amssymb}

\newcommand{\vect}[1]{\mathbf{\boldsymbol{#1}}} 
\newcommand{\matr}[1]{\mathbf{#1}} 

\newcommand{\calB}{\mathcal{B}}
\newcommand{\calPm}{\mathcal{P}^{m}}
\newcommand{\calDPm}{\mathcal{DP}^{m}}

\newcommand{\calWus}{\mathcal{W}^{\text{u/s}} }

\newcommand{\Xpol}{ \vect{X}_{\text{pol}} }

\newcommand{\Xus}{ \vect{X}^{\text{u/s}} }
\newcommand{\Xuspol}{ \Xpol^{\text{u/s}} }
\newcommand{\Bpol}{ \vect{B}_{\text{pol}} }

\newcommand{\Dpol}{ \vect{D}_{\text{pol}} }
\newcommand{\Xcyc}{ \vect{X}_{\text{cyc}} }

\newcommand{\rmd}{\mathrm{d}}

\newcommand{\hatn}{ \hat{ \vect{n} } }
\newcommand{\hats}{ \hat{ \vect{s} } }

\newcommand{\bN}{ \boldsymbol{N}  }

\usepackage{tikz}

\tikzset{down square arrow/.style={to path={-- ++(0,-.25) -| (\tikztotarget)}}}
\tikzset{up square arrow/.style={to path={-- ++(0,+.25) -| (\tikztotarget)}}}

\begin{document}

\preprint{APS/123-QED}

\title{On the shifts of stable and unstable manifolds\\ of a hyperbolic cycle under perturbation}

\author{Wenyin Wei$^{1,2,3}$}%
\author{Jiankun Hua$^{3,4}$}
\author{Alexander Knieps$^{3}$}%
\author{Yunfeng Liang$^{1,3,}$}
 \email{y.liang@fz-juelich.de}
\affiliation{%
$^1$ Institute of Plasma Physics, Hefei Institutes of Physical Science, Chinese Academy of Sciences, Hefei 230031, People's Republic of China
}%
\affiliation{
$^2$ University of Science and Technology of China, Hefei 230026, People's Republic of China
}%
\affiliation{
$^3$ Forschungszentrum J\"{u}lich GmbH, Institut f\"{u}r Energie- und Klimaforschung - Plasmaphysik, J\"{u}lich 52425, Germany
}%
\affiliation{
$^4$ International Joint Research Laboratory of Magnetic Confinement Fusion and Plasma Physics, State Key Laboratory of Advanced Electromagnetic Engineering and Technology, School of Electrical and Electronic Engineering, Huazhong University of Science and Technology, Wuhan 430074, People’s Republic of China
}%

\date{\today}

\begin{abstract}
Stable and unstable manifolds, originating from hyperbolic cycles, fundamentally characterize the behaviour of dynamical systems in chaotic regions. This letter demonstrates that their shifts under perturbation, crucial for chaos control, are computable with minimal effort using functional derivatives by considering the entire system as an argument. The shifts of homoclinic and heteroclinic orbits, as the intersections of these manifolds, are readily calculated by analyzing the movements of the intersection points.
\end{abstract}

\maketitle


\paragraph{Introduction.---} Building on the theory developed to describe the shifts of orbits and periodic orbits under perturbation~\cite{wei2024orbitshifts}, we analyze the shifts of stable and unstable manifolds of a hyperbolic cycle to gain a more predictive understanding of the global structure of a system and its behaviour under perturbation. This theory is motivated by the curiosity about strike line splitting observed in experiments \cite{yunfeng2013} and the complex intertwining manifold structure found in simulations \cite{frerichs2010, frerichs2015, frerichs2020} for magnetic confinement fusion (MCF). Since it does not require a specific system dimension, it has broad applications in fields governed by ordinary differential equations~(ODEs).

In the MCF community, the chaotic field formed by the distinctive structure of stable and unstable manifolds has been extensively studied due to the practical need for ELM (edge localized mode) mitigation and suppression in tokamaks \cite{yunfeng2007, yunfeng2010, yunfeng2013}. The process involves introducing external resonant magnetic perturbations (RMPs) to enable island chains to grow and corrode each other, creating a chaotic field \cite{howard1984, abdullaev2015, abdullaev2016magnetic}. The more resonant and stronger the perturbation to the helical field lines on the rational flux surface, the larger the corresponding island chain can become. The chaotic field is expected to reduce the transient heat flux released during ELMs by enhancing radial transport at the plasma edge. The magnetic spectrum analysis used in the past severely relies on the flux coordinates $(r, \theta, \varphi)$, which can lead to overlooking the probably largest chaotic field interwoven by the manifolds of the outermost X-cycle(s) due to the intrinsic singularity of these coordinates at the plasma edge $r=1$.

Functional derivatives, borrowed from functional analysis, are powerful tools that allow for the calculation of shifts of well-defined objects under system-wide perturbations, such as orbits, X/O-cycles, and invariant manifolds. This enables direct calculation of stable and unstable manifolds in real-world cylindrical coordinates, avoiding errors from transforming cylindrical to flux coordinates, which are unavailable in chaotic fields due to the requirement for strict system integrability. The stable and unstable manifolds, invariant by definition (all trajectories starting within a manifold remain inside it), thereby reveal the primary motion pattern of the system with the distinctive tangle structure. Because the edge chaotic field of stellarators can be enlarged significantly by the plasma response as the volume-averaged $\beta$ increases \cite{hudson2002, xu2023, zhou2022}, making it crucial to retain the ability to adjust island divertor for stellarators during operation.

The presented theory has extensive applications across various research domains concerned with the structure and bifurcation behaviour of dynamical systems \cite{meiss2015, harsoula2018, naplekov2021, kuznetsov2019, falessi2015, digiannatale2018, pegoraro2019} and the shape of invariant manifolds \cite{opreni2022, haro2016}, such as aircraft orbit optimization \cite{trelat2012, runqi2019}. The island-around-island hierarchy in near-integrable Hamiltonian systems has received considerable attention \cite{meiss1986, alus2014}, where remnant islands exist in the gaps between stable and unstable manifolds. The theory described here is not limited by the system's number of dimensions and does not require the field to be divergence-free as a magnetic field, thereby harvesting broad applicability.

\paragraph{Deduction and demonstration.---} For arbitrary $N$-dimensional dynamical systems $\dot{  \vect{x}  } = \vect{B}(\vect{x}) $ and 3D toroidal vector fields $\dot{  \vect{x} }_{\text{pol}} =
R \vect{B}_{\text{pol}} / B_\phi ~ (\vect{x}_{\text{pol}}, \phi) $, 
write a trajectory \textit{resp.} as $\vect{X}(\vect{x}_0, t)$ and $\Xpol(\vect{x}_{0,\text{pol}}, \phi_s, \phi_e )$, where $\vect{x}_{0}$ and $\vect{x}_{0,\text{pol}}$ are initial conditions, $\vect{B}_{\text{pol}}$ and $B_\phi$ are poloidal and toroidal components of the field in cylindrical coordinates, $\phi$ the azimuthal angle, $\phi_s$ and $\phi_e$ the starting and ending angles.
\begin{figure*}
\includegraphics[width=\linewidth]{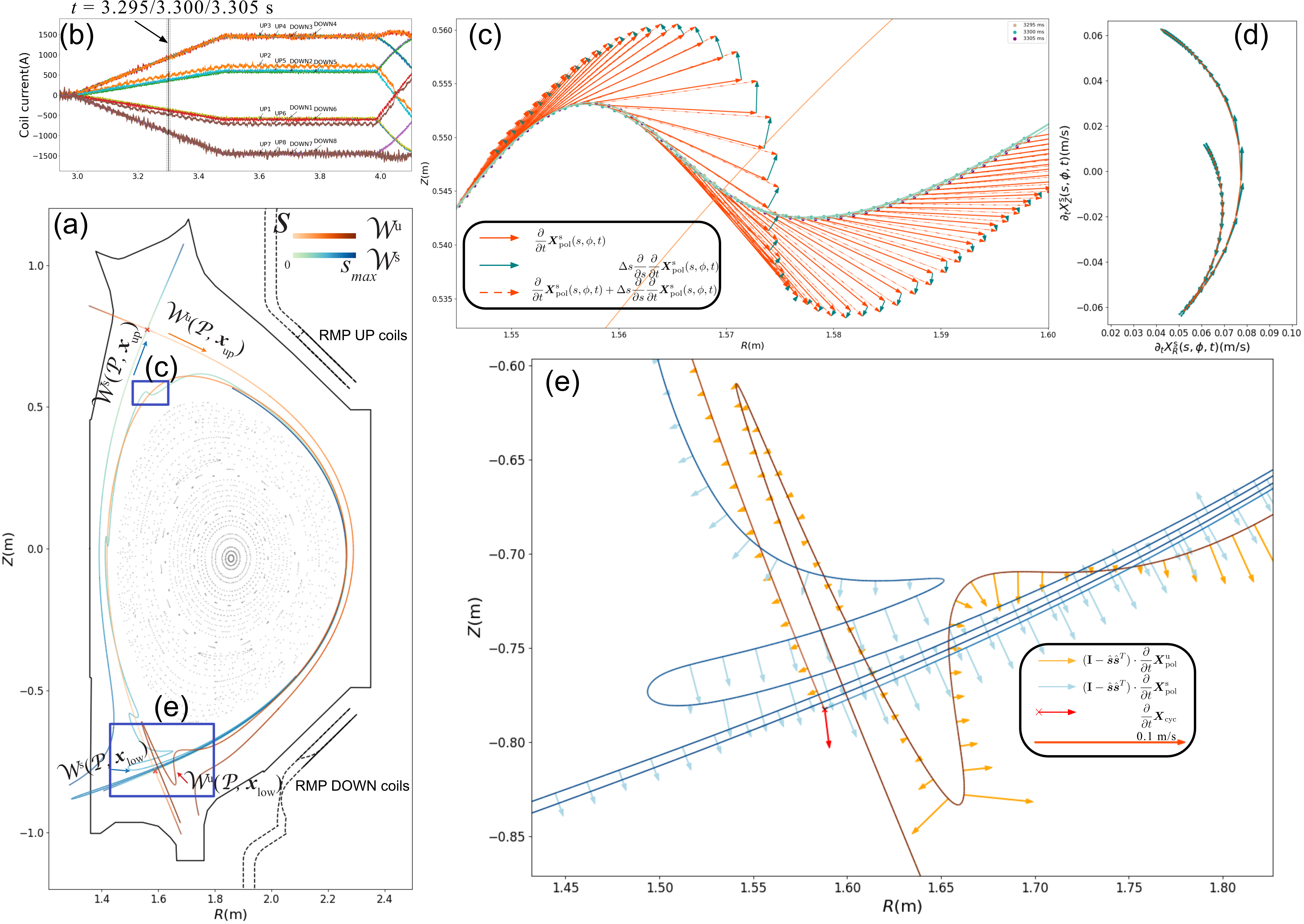}
\caption{\label{fig:EAST_103950_manifold_shifts} (a) Poincar\'e plot for EAST \#103950 shot at $t=3.300$ s and $\phi=0$ rad. $\pm 5$ ms and $\pm \pi/100$ rad data are used to implement the central difference scheme to compute the temporal derivative $\partial_{t}$ and $\phi$-derivative $\partial_\phi$. $\partial_t \calB$ is taken to be the perturbation field $\Delta\calB$ so the formula~\eqref{eq:manifold_growth_frist_variation} shall describe the progression of the manifold shift velocity $\partial_t \Xuspol(s,\phi,t)$ along the arc of $\calWus(\calPm_\phi, \Xcyc(\phi)~)$. (b) Time series of RMP coil currents. (c) The shift velocity $\partial_t \Xpol^{\text{s}} (s,\phi,t)$ of a branch of the stable manifold $\mathcal{W}^{\text{s}} (\mathcal{P}, \vect{x}_{\text{low}})$. $\Xpol^{\text{s}} (s,\phi,t)$ at the same $s$ but different $t$ are marked with different colors, by which the shift velocity is calculated.  (d) The orange arrow representing the shift velocities $\partial_t \Xpol^{\text{s}}(s,\phi,t)$ in (c) are collected to be an orange curve, while the teal arrows representing $\Delta s ~ \partial_s \partial_t  \Xpol^{\text{s}}(s,\phi,t)$ in (c) are also drawn to show an expected end-to-end tendency consistent with the orange curve.  (e) Perpendicular shift velocities of manifolds and the lower X-point drawn in the divertor region.}
\end{figure*}
In this letter, 3-D systems are focused due to their ubiquitous existence in nature. However, most equations do not need to be restricted to 3-D, of which the necessary extension will be pointed out when appropriate. 

The progression of first variations along a trajectory obey~\cite{wei2024orbitshifts}
\begin{subequations}
\begin{align}
& \frac{ \partial }{ \partial t } 
\delta \vect{X}
[\calB;\Delta\calB]
( \vect{x}_{0}, t )
=
\matr{A}~\delta\vect{X}
+
\Delta\vect{B} ,
\label{eq:deltaX_timeforward}
\\
&
\frac{ \partial }{ \partial \phi_e } 
\delta \Xpol
[\calB;\Delta\calB]
( \vect{x}_{0,\text{pol}}, \phi_s, \phi_e )
=
\matr{A}~\delta\Xpol 
+
\Delta \frac{R\Bpol }{B_\phi},
\label{eq:deltaXpol_phiforward}
\end{align}
\end{subequations}
where $\matr{A}= \matr{A}(\Xpol, \phi_e) := \partial_{(R,Z)} (R \Bpol / B_\phi)|_{(\Xpol,\phi_e)} $, $\Delta (R\Bpol / B_\phi)$ is short for $(\Delta\calB\cdot\rmd / \rmd \calB)  (R \Bpol / B_\phi)$, that is $R\Delta\Bpol / B_\phi - (R\Bpol / B_\phi^2) ~ \Delta B_\phi$. Since in this letter it is always taken into account only one perturbation $\Delta\calB$ instead of a series of different ones $\{ \Delta \calB_i \}$, so the functional arguments inside the square bracket of $\delta^k \vect{X} [\calB; \Delta\calB,\dots, \Delta\calB]$ can be omitted, simply written $\delta^k \vect{X}$.

Imposing $\Delta\calB\cdot \rmd / \rmd \calB$ on both sides of Eqs.~\eqref{eq:deltaX_timeforward} and \eqref{eq:deltaXpol_phiforward} gives the high-order ODEs that can be integrated in time to acquire $\delta^k \vect{X}$ and $\delta^k \vect{X}_\text{pol}$, of which the technique has been shown in \cite{wei2024orbitshifts}. The difference with its application in this letter is that in \cite{wei2024orbitshifts} the initial condition was fixed for orbits or bound to the ending point for periodic orbits. In this letter, to study the shifts of stable and unstable manifolds under perturbation, one needs to anchor on the manifold a series of points whose trajectories can fill the manifold,  and estimate the shifts of these points as the initial conditions for the first variation progression equations.

To determine the initial conditions, Taylor expand $\Xuspol(s,\phi)$ near the X-cycle. Denote the X-cycle R-Z coordinates at $\phi$ by $\Xcyc(\phi)$, whose variations under perturbation, $\delta^k\Xcyc$, have been deduced to be \cite{wei2024orbitshifts}.
\begin{align}
\Xuspol(s,\phi) &= \Xuspol(0,\phi) + \vect{v}^{u/s}\cdot s + \mathcal{O}(\|s\|^2) ,\\
\delta\Xuspol(s,\phi) &= \delta \Xuspol(0,\phi) + \delta\vect{v}^{u/s}\cdot s + \mathcal{O}(\|s\|^2) 
\nonumber \\
& \text{(omit u/s and use $^{\prime}$ for $\Delta\calB\cdot \rmd / \rmd \calB$) }  
\nonumber \\
 & = \delta \Xcyc(\phi) + \matr{R} \vect{v}\theta^\prime \cdot s + \mathcal{O}(\|s\|^2) , \\
 \delta^2 \Xuspol(s,\phi) 
 &= \delta^2 \Xcyc(\phi) 
    \nonumber\\
 &+ (\matr{R}^2 \vect{v} \theta^{\prime 2}+\matr{R} \vect{v} \theta^{\prime\prime} ) \cdot s+  + \mathcal{O}(\|s\|^2) ,
\end{align}
where $\vect{v}=[\cos\theta, \sin\theta]^\text{T} $ denotes the eigenvector of $\calDPm_\phi (\Xcyc(\phi))$ corresponding to the grown manifold, $\calPm_\phi$ means the Poincar\'e map at $\phi$-section. The variations of $\theta$ under perturbation need to be calculated with knowledge of $(\calDPm_\phi(\Xcyc(\phi)) )^\prime$, which has been concluded in \cite{wei2024orbitshifts}. How the eigenvalues and eigenvectors change as a matrix varies is analysed in Supplemental Material~\cite{supp}.


On the other hand, the \emph{invariant manifold formula in cylindrical coordinates}~(Eq.~(16) in \cite{wei2023}) describes the growth direction for the 1-D submanifold $\calWus(\calPm_\phi, \Xcyc(\phi))$ on the $\phi$-section. The formula is repeated below,
\begin{align}
\partial_s \Xuspol (s,\phi)
= 
\frac{
    R\Bpol / B_\phi
    - \partial_\phi \Xuspol
}{
    \pm  \|\cdots \|
}
,
\label{eq:invariant_manifold_growth} 
\end{align}
where the $\dots$ in the denominator denotes the numerator, while the numerator is denoted as $\Dpol$ from now on. With $\Delta\calB\cdot \rmd / \rmd \calB$ exerted on both sides, the formula shall give $\delta \partial_s \Xuspol$ as result, which depicts the progression of $\delta\Xus(s,\phi)$ along the arc of $\calWus (\calPm, \Xcyc(\phi)) $.
\begin{align}
\delta \frac{\partial }{\partial s }  \Xuspol
&=
(\frac{\rmd s}{\rmd \phi})^{-1}
\Dpol^{\prime} - 
(\frac{\rmd s}{\rmd \phi})^{-3}
\Dpol\cdot \Dpol^{\prime}   \Dpol 
    ,
\label{eq:manifold_growth_frist_variation}
\end{align}
where $\rmd s / \rmd \phi := \pm \| \Dpol \|$, that is the derivative of the arc length $s$ of $\calWus(\mathcal{B}, \gamma)$  \textit{w.r.t.} $\phi$ along the local field line (also the denominator of RHS of the manifold growth formula).
The detailed deduction and extension to higher orders are put in supplemental material \cite{supp}. EAST tokamak is taken as an example to demonstrate how to use the Eq.~\eqref{eq:manifold_growth_frist_variation} as shown in Fig.~\ref{fig:EAST_103950_manifold_shifts}.

This way is hard to extend from 3-D to arbitrary $N$-D systems, since it requires a cylindrical-like coordinate system to have an azimuthal angle-like coordinate to parameterize the manifold of concern. Another much extensible way is to focus on a 1-D submanifold of the (un)stable manifold of a map and parameterize it with the arc length $s$ along the submanifold. By this definition,
\begin{align}
\Xuspol (s+\Delta s,\phi) := \calPm(\Xuspol(s,\phi),\phi),
\end{align}
where $\Delta s = (s, \phi)$ denotes the arc length change induced by once mapping. Differentiate both sides of it in $s$, then
\begin{align}
(1 + \partial_s \Delta s )
&\partial_{s} \Xuspol (s + \Delta s, \phi ) \nonumber \\
=\calDPm &( \Xuspol(s,\phi) ) 
\cdot 
\partial_{s} \Xuspol(s,\phi).
\label{eq:pXps_along1dW}
\end{align}
Then, fix $\phi$ so that it can be omitted from the argument list of $\Delta s(s,\phi)$ and $\Xuspol(s,\phi)$. Remove the subscript $_{\text{pol}}$ to consider a general map $\mathcal{P}^m:~\mathbb{R}^N \to \mathbb{R}^N$ rather than necessarily a Poincar\'e map, thereafter
\begin{align}
(1 + \partial_s \Delta s )
\partial_{s} & \Xus (s + \Delta s )  
    \nonumber \\
=\calDPm ( & \Xus(s) ) 
\cdot 
\partial_{s} \Xus(s),  
\end{align}
where we still use $\partial_s $ instead of $\rmd / \rmd s$ for the derivative of $\Delta s$ even though it is a univariate function now, because later we will vary $\calB$, which will be added to the argument list. Impose $\Delta \calB \cdot \rmd / \rmd \calB$ on both sides of the above equation, then
\begin{align}
& \underline{\delta \partial_s \Delta s} ~
\partial_{s}  \Xus (s + \Delta s )  
    \nonumber \\
&
+
(1 + \partial_s \Delta s ) 
\left(
    \underline{ \delta \partial_{s} \Xus (s + \Delta s )  }
    + \delta \Delta s ~ \partial_{s}^{2} \Xus (s + \Delta s )  
\right)
    \nonumber \\
&=
\Big[
    \delta \calDPm ( \Xus ) 
    + (\delta\Xus \cdot \mathcal{D}) \calDPm ( \Xus ) 
\Big]\cdot 
    \partial_{s} \Xus  
    \nonumber \\
&+ 
    \calDPm ( \Xus ) 
    \cdot 
    \delta \partial_{s} \Xus,
\end{align}
where $\delta\partial_s \Delta s = \delta\partial_s \Delta s [\calB; \Delta\calB](s,\phi)$. The underlined terms $\delta\partial_s \Delta s$ and $\delta \partial_s \Xus$ are needed to be integrated in $s$ to give $\delta\Delta(s)$ and $\delta \Xus (s)$. However, notice that the equation determines $N$ degrees of freedom~(DOFs) but the two underlined terms have $N+1$ DOFs~($\delta \partial_s \Xus$ is $N$-D and $\delta\partial_s \Delta s$ is 1-D). A scalar equation suffices to fill the missing 1 DOF. The equation comes from the definition of arc length, that is $|\partial_s \Xus|=1$. Detailed deduction and extension to higher orders are put in Supplemental Material~\cite{supp}. This approach is much more extensible since a high-dimensional (un)stable manifold can be decomposed into a set of 1-D (un)stable submanifolds, each of which is still invariant.

\paragraph{Remark.---}
The two variations $\delta\Xuspol(\vect{x}_{0, \text{pol}},\phi_s,\phi_e)$ and $\delta\Xuspol(s,\phi)$ do not necessarily coincide at the same point, but their perpendicular components do. From a perspective of differential geometry, it is their perpendicular component that plays the role of an intrinsic property, which is independent of how the manifold is parameterized, \textit{i.e.} how the manifold is embedded. Therefore, the argument list $(s,\phi)$ can be freely omitted for $\delta_{\perp}\Xuspol$ but not for $\delta_{\perp}\Xuspol$. $\delta\Xuspol$ given by Eq.~\eqref{eq:deltaXpol_phiforward} in fact parameterizes the manifold with $(s_0, \phi_e)$, where $s_0$ is the distance of the initial point to the X-cycle, and $\phi_e$ is the azimuthal angle, accounting for the number of turns the trajectory has completed ($\phi_e\in\mathbb{R}$ not just $[0,2 m \pi)$ ). Similarly, the argument list $(s_0, \phi_e)$ can also be dropped if there is no confusion.

The perpendicular components
\begin{align}
 \delta_{\perp}\Xuspol
&:= \hats_{\perp}\hats_{\perp}^\text{T} \cdot\delta\Xuspol, \\   
\delta_{\perp}\Xus
&:= \hatn \hatn^\text{T} \cdot \delta\Xus,
\end{align}
where $\hats_\perp := \matr{R}\hats $,  $\matr{R}:= [ 0, -1; 1, 0 ] $ the counterclockwise $90^\circ$ rotation matrix,  $\hats := \partial_{s} \Xuspol $ and $\hatn \equiv \hats \times \vect{B} / |\hats \times \vect{B}|$. To transform between $\delta_{\perp} \Xus$ and $\delta_{\perp}\Xuspol$, one simply needs to utilize the projection relationship $\delta_{\perp}\Xuspol  = \hats_\perp \hats_\perp^\text{T} \cdot \delta_{\perp} \Xus$. Notice that $\hats_\perp^\text{T}\cdot\hatn =  \cos\angle(\hats_\perp, \hatn)$.
The variations of the homo- and heteroclinic orbits/trajectories --- the intersection of stable and unstable manifolds --- are computable with the above perpendicular components of variations. Remember that the intersection point shift velocity obey
\begin{align}
\hats_{\perp}^{\text{u}} \hats_{\perp}^{\text{u}\text{T} }
\cdot
\delta \vect{X}^{\times}_{\text{pol}}  
&= \delta_{\perp} \vect{X}^{\text{u}}_{\text{pol}} ,
    \nonumber \\
\hats_{\perp}^{\text{s}} \hats_{\perp}^{\text{s}\text{T} }
\cdot
\delta \vect{X}^{\times}_{\text{pol}}  
&= \delta_{\perp} \vect{X}^{\text{s}}_{\text{pol}}  .
    \nonumber \\
\intertext{
The matrices $\hats_{\perp}^{\text{u}} \hats_{\perp}^{\text{u}\text{T} }$ and $\hats_{\perp}^{\text{s}} \hats_{\perp}^{\text{s}\text{T} }$ are not invertible alone, but their linear combination is. Therefore, }
\delta \vect{X}^{\times}_{\text{pol}}  
= (\hats_{\perp}^{\text{u}} \hats_{\perp}^{\text{u}\text{T} }
+ 
& \hats_{\perp}^{\text{s}} \hats_{\perp}^{\text{s}\text{T} }
)^{-1} 
\cdot 
\delta_{\perp} 
(\vect{X}^{\text{u}}_{\text{pol}} 
+ \vect{X}^{\text{s}}_{\text{pol}} )
\end{align}
Similarly, for the 2-D manifolds in a 3-D flow,
\begin{align}
\delta \vect{X}^{\times}  
= & (\hatn^{\text{u
}} \hatn^{\text{u}\text{T} }
+ 
\hatn^{\text{s}} \hatn^{\text{s}\text{T} }
)^{-1} 
\cdot 
\delta_{\perp} 
(\vect{X}^{\text{u}}
+ \vect{X}^{\text{s}} )
\end{align}
A (un)stable manifold in a higher-dimensional system may have normal vector spaces $\bN_{p} \calWus $ ($p\in \calWus$) of more than one dimension. In such cases, the above equations are complicated into 
\begin{align}
(\sum \hatn^{\text{u}}_i \hatn^{\text{u}\text{T} }_i )
\cdot
\delta \vect{X}^{\times}
&= \delta_{\perp} \vect{X}^{\text{u}} ,
\nonumber \\
(\sum \hatn^{\text{s}}_i \hatn^{\text{s}\text{T} }_i )
\cdot
\delta \vect{X}^{\times}
&= \delta_{\perp} \vect{X}^{\text{s}} ,
\nonumber \\
\delta \vect{X}^{\times}  
=  (\sum \hatn^{\text{u}}_i \hatn^{\text{u}\text{T} }_i  + & \sum \hatn^{\text{s}}_i \hatn^{\text{s}\text{T} }_i
)^{-1} 
\cdot 
\delta_{\perp} 
(\vect{X}^{\text{u}}
+ \vect{X}^{\text{s}} ),
\end{align}
where $\{ \hatn_{i}^{\text{u/s}} \}$ can be considered an orthnormal basis for $\bN_{p} \calWus$. For higher-order variations $\delta^k_\perp \vect{X}^{\times}$, the above equations do not need to make substantial changes.

With the above variations, one can conveniently calculate the variation of the quantities concerned, \textit{e.g.} the magnetic flux $\Phi := \iint_{S} B_{\phi}  \rmd S $ through a lobe $S$ surrounded by $\mathcal{W}^{\text{u}}$ and $\mathcal{W}^{\text{s}}$. Then, $
\delta \Phi = \iint_{S} \Delta B_\phi \rmd S  
+ \oint_{\partial S} B_\phi ~ \delta_{\perp} \Xuspol \cdot \hats_\perp ~ \rmd l$,
where the normal vector $\hats_\perp$ is chosen to be outwards for the lobe area $S$. The first variation of the lobe area is simply $
\delta A = \oint_{\partial S} \delta_{\perp} \Xuspol \cdot \hats_\perp ~ \rmd l.
$

Given the importance of the perpendicular component of variations as an intrinsic property, one may wish to have an equation as concise as Eq.~\eqref{eq:deltaXpol_phiforward} to describe the perpendicular component of a variation, such that the parallel component does not need to be solved. This turns out to be impossible due to other terms than $\matr{A}\cdot \delta_{\perp} \Xpol + \Delta_{\perp} 
\left(
R\Bpol / B_{\phi}
\right) $ 
in
\begin{align}
\frac{\partial}{\partial \phi} ~ \delta_{\perp} \Xpol
& = \frac{\rmd}{\rmd \phi}
\left( 
(\hats_{\perp} ~ \hats_{\perp}^\text{T})\cdot \delta \Xpol 
\right)
    \\
= \matr{A}\cdot \delta_{\perp} \Xpol &+ \Delta_{\perp} 
\frac{R\Bpol}{B_{\phi}}
+ 
    (\frac{\rmd}{\rmd \phi} \hats_{\perp} ~ \hats_{\perp}^\text{T}
    +\hats_{\perp} ~ \frac{\rmd}{\rmd \phi} \hats_{\perp}^\text{T}) \cdot \delta \Xpol, \nonumber
\end{align}
where $\rmd / \rmd \phi$ is the total derivative \textit{w.r.t.} $\phi$ along the field line, and $\Delta_\perp (R\Bpol / B_\phi) := \hats_{\perp} \hats_{\perp}^\text{T}\cdot \Delta (R\Bpol / B_\phi) $.

\paragraph{Conclusion and discussion.---}
The lobe structure of the stable and unstable manifolds of the outermost X-cycles in an MCF machine can damage fragile components if the plasma response drives the lobes to unexpected locations~\cite{zhou2022,knieps2022plasma,knieps2022anisotropic}. The heat flux distribution on the divertor plates depends on the inter-shading order of the lobes. Without knowledge of the shapes of invariant manifolds, one may be puzzled by the peculiar ribbon pattern of the deepest field line incursion into the plasma $\rho_{\text{min}}$~\cite{jia2021,yunfeng2013, frerichs2010, frerichs2020}. The capacity of a divertor to dissipate the power flowing into the scrape-off layer can be enhanced rather than diminished due to unconsciously threatening fragile components if the magnetic topology is well understood.

A better understanding of the chaotic field enables a fusion machine to avoid relying solely on plasma diffusion to increase the power decay length $\lambda_{q}$ of the scrape-off layer~\cite{eich2011}. Instead, it can be enhanced through proactive chaotic field stimulation to disperse heat flux in midair before it reaches the target plate. This understanding can also aid in controlling the behaviour of complex systems in other research domains. Agile and accurate predictions of the shifts under perturbation can guide the swift identification of the desired perturbations to the system.

\begin{acknowledgments}
This work was supported by National Magnetic Confined Fusion Energy R\&D Program of China (No. 2022YFE03030001) and National Natural Science Foundation of China (Nos. 12275310 and 12175277). 
\end{acknowledgments}

\appendix


\bibliography{apssamp}

\end{document}